\documentclass[aps,pra,twocolumn,amsmath,amssymb,showpacs,superscriptaddress,notitlepage,longbibliography]{revtex4-1}
\usepackage[colorlinks=true,linkcolor=blue,anchorcolor=red,citecolor=blue,urlcolor=blue]{hyperref}
\usepackage{bm}
\usepackage{graphicx}
\usepackage{color}
\usepackage{multirow}
\usepackage{amsmath}
\usepackage{simplewick}
\begin{document}
%\begin{CJK*}{GBK}{song}
%\begin{CJK*}{UTF8}{gbsn}

\title{Universal relations for hybridized $s$- and $p$-wave interactions from spin-orbital coupling}

%\author{Fang Qin (ñû•P)}
\author{Fang Qin}
\affiliation{Shenzhen Institute for Quantum Science and Engineering and Department of Physics, Southern University of Science and Technology (SUSTech), Shenzhen 518055, China}
\affiliation{CAS Key Laboratory of Quantum Information, University of Science and Technology of China, Chinese Academy of Sciences, Hefei, Anhui 230026, China}

\author{Pengfei Zhang}
\email{pengfeizhang.physics@gmail.com}
\affiliation{Walter Burke Institute for Theoretical Physics, California Institute of Technology, Pasadena, California
91125, USA}
\affiliation{Institute for Quantum Information and Matter, California Institute of Technology, Pasadena, California
91125, USA}

\date{\today}

\begin{abstract}
In this work, we study the universal relations for one-dimensional spin-orbital-coupled fermions near both $s$- and $p$-wave resonances using effective field theory. Since the spin-orbital coupling mixes different partial waves, a contact matrix is introduced to capture the non-trivial correlation between dimers. We find the signature of the spin-orbital coupling appears at the leading order for the off-diagonal components of the momentum distribution matrix, which is proportional to $1/q^{3}$ ($q$ is the relative momentum). We further derive the large frequency behavior of the Raman spectroscopy, which serves as an independent measurable quantity for contacts. Finally, we give an explicit example of contacts by considering a two-body problem.
\end{abstract}

\maketitle
%\newpage%\clearpage
%\end{CJK*}

\section{Introduction}\label{1}
In ultracold atomic gases, a series of universal relations was established to set up a bridge between the short distance two-body correlations and the macroscopic thermodynamic properties~\cite{Tan20081,Tan20082,Tan20083,Braaten20081,Braaten20082,Zhang2009,Platter2016}.
These relations are connected by a set of key parameters called the
contacts that have already been examined in experiments~\cite{Exp2010,Exp2012,Exp2013,Exp2016,Exp2019}.
Later, the universal relations were also studied in higher partial-wave systems~\cite{YuP2015,YuP2015Erratum,YoshidaP2015,HeP2016,PengP2016,ZhangD2017}, low-dimensional systems~\cite{Barth1D2011,Cui1D20161,Cui1D20162,Ovidiu1D2017,Yin1D2018,Valiente2D2011,Werner2D20121,Werner2D20122,Hofmann2D2012,Zhang2D2017,Peng2D2019}, laser-dressed systems~\cite{Qin20181,Qin20182}, and were taken into account in three-body correlations~\cite{Braaten3b2011,Braaten3b2014,Fletcher3b2017,Zhang3b2D2017,Zhang3banyD2017}.

Recent experimental realization of the spin-orbital coupling (SOC) in ultracold gases~\cite{ExpSOC2011,ExpSOC20121,ExpSOC20122,ExpSOC20123,ExpSOC2013} also leads to interesting few- and many-body physics~\cite{GoldmanSOC2014,ZhangSOC2014,ZhaiSOC2012,ZhaiSOC2015,Bethe-Peierls2012,Bethe-Peierls2013,Zhenhua2012,Zhenhua2013,Cui2012pseudopotential,Cui2017,Dong2013njp,Dong2014pra,Dong2015Atoms,Dong2015nc,Yong-Chang2015prl,Shang-Shun2016pra,Zhu2016jpb}.
In particular, the universal relations for the spin-orbital coupled Fermi gases attract much attention~\cite{PengSOC2018,JieSOC2018,ZhangSOC2018,PengSOC2020,QinSOC20201}. Since the SOC breaks the rotational symmetry, it would mix different partial waves at the two-body level. It is interesting to study the universal relations for systems with one-dimensional (1D) SOC with both $s$- and $p$-wave interactions. Experimentally, a system with overlapping resonances of $s$ and $p$ waves has been realized in $^{40}$K atoms using the optical control \cite{PengSP2018}, where, in principle, additional SOC can be engineered directly.

Motivated by these developments, in this work, we study the universal relations for a 1D Fermi gas with hybridized $s$- and $p$-wave interactions from SOC. Importantly, we find that the $q^{-3}$ tail in the spin-mixing (off-diagonal) terms of the momentum distribution matrix is a direct manifestation of the SOC-induced strong interplay of $s$- and $p$-wave interactions, which can be observed through time-of-flight measurement. Further, we study the Raman spectroscopy and also find that the spin-mixing term of the Raman spectroscopy matrix is a useful experimental probe that can be used to detect the hybridization of $s$- and $p$-wave interactions. In the end, we calculate the contacts in two-body bound states as an explicit example of the contact matrix \cite{YoshidaTensor2016,ZhangMatrix2017} in the hybridized $s$- and $p$-wave Fermi gases. It is found that there is a peak for the two-body hybridized contact of the $s$  and $p$ waves near the degenerate point of $s$- and $p$-wave scattering lengths, indicating a strong interplay between $s$- and $p$-wave dimers as expected.

{The paper is organized as follows: In Sec.~\ref{2}, we give the model Hamiltonian and calculate the two-body physics.
In Sec.~\ref{3}, we give the definition of the contacts. We calculate the large-momentum distribution tail in Sec.~\ref{4} and we calculate the high-frequency tail of the Raman spectroscopy in Sec.~\ref{5}.
In addition, we discuss other universal relations in Sec.~\ref{6}.
As a concrete example, we calculate the contacts in two-body states in Sec.~\ref{7}.
Finally, we provide a brief summary and discussions in Sec.~\ref{8}.}

\section{Model}\label{2}
The experiment~\cite{PengSP2018} shows that the optical control of a $p$-wave magnetic Feshbach resonance can realize the noninteracting state between spin-down atoms near $s$-wave resonance, based on a laser-field-coupled bound-to-bound transition between the $p$-wave closed-channel molecular states. It can also be used to shift the $p$-wave Feshbach resonance associated with the spin-up atoms close to the resonance of the $s$ wave in $^{40}$K atoms. We consider a fermion system with an $s$-wave interaction between atoms with spin $\uparrow$ and $\downarrow$, together with a $p$-wave interaction between two spin-$\uparrow$ fermions. Without SOC, the interesting few- and many-body physics have been studied in \cite{HuSP2016,YangSP2016,JiangSP2016,QinSP2016,ZhouSP2017}.  After adding the SOC, the effective 1D Lagrangian is given by ($\hbar=1$ throughout the paper)
{\begin{equation}\label{eq:Lagrangian0}
\begin{aligned}
\hat{L}
=&\sum_{k}\Psi_{k}^{\dag}\left(i\partial_{t} - {\cal H}_{k}^0 \right)\Psi_{k}^{}\\
&- \frac{g_{S}^{}}{L}\sum_{Q,k,k'}\psi_{Q/2-k',\downarrow}^{\dagger}\psi_{Q/2+k',\uparrow}^{\dagger}\psi_{Q/2+k,\uparrow}^{}\psi_{Q/2-k,\downarrow}^{}\\
&- \frac{g_{P}^{}}{4L}\sum_{Q,k,k'}k'\psi_{Q/2-k',\uparrow}^{\dagger}\psi_{Q/2+k',\uparrow}^{\dagger}k\psi_{Q/2+k,\uparrow}^{}\psi_{Q/2-k,\uparrow}^{},
\end{aligned}
\end{equation}  where $L$ is the system size and $g_{S}^{}$ ($g_{P}^{}/4$) is the effective 1D $s$($p$)-wave coupling constant.} We have defined $\Psi_{k}=(\psi_{k,\uparrow}^{},\psi_{k,\downarrow}^{})^{T}$, where $\psi_{k,\sigma}^{}$ is the field operator for the fermionic atoms with momentum $k$ and spin $\sigma$. The single-particle Hamiltonian is {${\cal H}_{k}^0=\frac{(k I_2+k_{0}\sigma_{z})^{2}}{2m}+\Omega\sigma_{x}$}, where atoms in the state $\left|\uparrow\right>$ are coupled to the state $\left|\downarrow\right>$ by the Raman laser with the strength $\Omega$, and $2k_{0}$ is the momentum transfer during the two-photon processes. {Here, $\sigma_{x/y/z}$ is the Pauli matrix and $I_2$ is the $2\times 2$ identity matrix.}

Before performing calculations, we would like to comment on the validity of the the Lagrangian \eqref{eq:Lagrangian0}. Similar to previous studies~\cite{1DHamiltonian2014,1DHamiltonian2013,1DHamiltonian2018}, the microscopic Hamiltonian in three dimensions can be divided into three parts: $H=H_{0}+H_{\perp}+H_{\text{int}},$ where $H_{0}$ is the free Hamiltonian with SOC along the $z$ direction, $H_{\text{int}}$ is the three-dimensional (3D) interacting Hamiltonian with the 3D scattering parameters, and $H_{\perp}$ contains the transverse kinetic energy and transverse confinement potential. In real experiments, the trapping frequency of the transverse confinement potential is $10^5$ Hz \cite{Liao2010}, which is much larger than a moderate SOC strength of $\sim 10^3$ Hz \cite{ExpSOC20121}. Since the length of the SOC is much longer than the potential range, i.e., the SOC in experiments~\cite{ExpSOC2011,ExpSOC20121,ExpSOC20122,ExpSOC20123,ExpSOC2013} can barely reach the very short-range regime of the very deep short-range potential~\cite{Cui2017}, the SOC will not modify the scattering inside the short-range potential. Consequently, when solving the scattering problem, we could separate the $z$ coordinate into regions with $z\lesssim 1/\sqrt{\omega_\perp m}$ and $z\gtrsim 1/\sqrt{\omega_\perp m}$. In the region of $z\lesssim 1/\sqrt{\omega_\perp m}$, the problem is intrinsically 3D at high energy $\sim~\omega_\perp$ and one could neglect both the kinetic energy in the $z$ direction as well as the SOC. This gives the wave function at $z\sim 1/\sqrt{\omega_\perp m}$ up to leading order. The higher-order corrections are proportional to $E_{z}/(\omega_\perp m)$, which is sufficiently small compared with the leading-order term, where $E_{z}$ can be the kinetic energy in the $z$ direction or the SOC strength. The wave function for $z\gtrsim 1/\sqrt{\omega_\perp m}$ is determined by matching the boundary condition at $z\sim 1/\sqrt{\omega_\perp m}$, which can be modeled by a contact pseudopotential. Since, to the leading order, the boundary condition is determined by a Hamiltonian without SOC, there is no coupling between the $s$- and the $p$-wave contact pseudopotential, which leads to our Lagrangian \eqref{eq:Lagrangian0}. This analysis is consistent with the results presented in Refs.~\cite{1DHamiltonian2014,1DHamiltonian2013,1DHamiltonian2018} when the transverse trapping frequency $\omega_{\perp}$ of the confinement potential is much larger than the strength of the SOC.

{To conveniently calculate the Feynman diagrams, the above Lagrangian (\ref{eq:Lagrangian0}) can be rewritten as follows: }
\begin{equation}\label{eq:Lagrangian}
\begin{aligned}
\hat{L}
=&\sum_{k}\Psi_{k}^{\dag}\left(i\partial_{t} - {\cal H}_{k}^0 \right)\Psi_{k}^{} + \sum_{Q;\alpha=S,P}\frac{\varphi^{\dagger}_{Q,\alpha}\varphi_{Q,\alpha}^{}}{g_{\alpha}^{}}\\
&- \frac{1}{2\sqrt{L}}\sum_{Q,k}\left[ \varphi_{Q,S}^{\dagger}\left(\Psi^{T}_{\frac{Q}{2}+k}{\sigma_{S}}\Psi_{\frac{Q}{2}-k}^{}\right) + \text{H.c.} \right]\\
&- \frac{1}{2\sqrt{L}}\sum_{Q,k}k\left[ \varphi_{Q,P}^{\dagger}\left(\Psi^{T}_{\frac{Q}{2}+k}{\sigma_{P}}\Psi_{\frac{Q}{2}-k}^{}\right) + \text{H.c.} \right],
\end{aligned}
\end{equation} {where we have used the definitions
$$\varphi^{\dagger}_{Q,S}\equiv g_{S}^{}\sum_{k'}\psi_{Q/2-k',\downarrow}^{\dagger}\psi_{Q/2+k',\uparrow}^{\dagger}/\sqrt{L}$$ and
$$\varphi^{\dagger}_{Q,P}\equiv \frac{1}{2}g_{P}^{}\sum_{k'}k'\psi_{Q/2-k',\uparrow}^{\dagger}\psi_{Q/2+k',\uparrow}^{\dagger}/\sqrt{L}.$$}
$\varphi_{Q,S}^{}$ ($\varphi_{Q,P}^{}$) is the field operator of the $s$($p$)-wave dimer with momentum $Q$. Note that although we have introduced a dimer field for convenience, the Lagrangian contains no dynamics of dimers and is essentially single channel. The generalization to two-channel models is straightforward and gives the same universal relations to the leading order. Interaction vertexes ${\sigma_{S}}$ and ${\sigma_{P}}$ can be related to Pauli matrices $\sigma_{j}$ as ${\sigma_{S}}=i\sigma_{y}$, ${\sigma_{P}}=\frac{1}{2}(1+\sigma_{z})$, which is equivalent to
\begin{align}
&\frac{1}{2}\Psi^{T}_{Q/2+k}{\sigma_{S}}\Psi_{Q/2-k}^{} = \psi_{Q/2+k,\uparrow}^{}\psi_{Q/2-k,\downarrow}^{},\\
&\Psi^{T}_{Q/2+k}{\sigma_{P}}\Psi_{Q/2-k}^{} = \psi_{Q/2+k,\uparrow}^{}\psi_{Q/2-k,\uparrow}^{}.
\end{align}

To regularize the possible divergence, we impose a momentum cutoff at $k\sim \Lambda$. The bare interaction parameters $g_{S}^{}$ and $g_{P}^{}$ can be related to the physical scattering lengths by
\begin{equation}\label{eq:renormalization}
a_{s}  = - \frac{2}{mg_{S}^{}},\ \ \ \ \ \frac{1}{a_{p}} = \frac{4}{mg_{P}^{}} + \frac{2\Lambda}{\pi},
\end{equation} where $a_{s}$ ($a_{p}$) is the 1D $s$($p$)-wave scattering length.

{According to Eq.~(\ref{eq:renormalization}), $g_{S}^{}$ has unit of length$^{-1}$ and $g_{P}^{}$ has unit of length. Moreover, since $\psi_{\sigma}(x)=\sum_{k}e^{ikx}\psi_{k,\sigma}/\sqrt{L}$, the dimension of $\psi_{k,\sigma}$  is length$^{0}$, knowing the dimension of $\psi_{\sigma}(x)$ is length$^{-1/2}$. In addition, $\sum_{k}\rightarrow L\int_{-\infty}^{\infty}dk/(2\pi)$ is  dimensionless. Therefore, the unit of $\varphi_{Q,S}^{}$ is length$^{-3/2}$, and the unit of $\varphi_{Q,P}^{}$ is length$^{-1/2}$. $\varphi_{Q,S}^{}$ and $\varphi_{Q,P}^{}$ have different scaling dimensions and these quantities differ by a factor of length. }

As mentioned before, we focus on a very special quasi-1D case where $\Omega$ and $k_0^2$ are much smaller than the transverse trapping frequency $\omega_{\perp}$ of the confinement potential, i.e., $\omega_{\perp} \gg\Omega$ and $\omega_{\perp} \gg k_0^2/m$.  Consequently, in this limit, the scattering length would not depend on the SOC parameters, i.e., the reduction from 3D to 1D of the interaction receives no contribution from the SOC, consistent with the previous references~\cite{1DHamiltonian2014,1DHamiltonian2013,1DHamiltonian2018}.
In this case, the quasi-1D $s(p)$-wave scattering length connected to the three-dimensional (3D) one  is given by~\cite{quasi1Ds1998,quasi1Ds2003,quasi1Ds2012,quasi1Ds2017,quasi1Dp2004,quasi1Dp2008,quasi1Dp2014,quasi1Dp2017,quasi1Dp2020}
\begin{equation}\label{eq:a1D}
a_{s}^{} = - \frac{\ell_{\perp}^{2}}{2a_{3D}} + \frac{{\cal C}\ell_{\perp}}{2}, ~~~~~~~~~
a_{p}^{} = \frac{3V_{p}}{\ell_{\perp}^{2}},
\end{equation} where $a_{3D}$ is the 3D $s$-wave scattering length, ${\cal C}=1.4603$, $\ell_{\perp}=\sqrt{2/(m\omega_{\perp})}$, $\omega_{\perp}$ is the transverse trapping frequency, and $V_{p}$ is  the 3D $p$-wave scattering volume.

%%%%%%%%%%%%%%%%%%%%%%%%%%%%%%%%%%%%%%%%%%%%%%%%%%
\begin{figure}
\includegraphics[width=8.5cm]{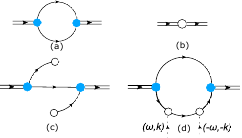}
\caption{(a) Diagrams for the matrix elements of the dimer-atom interaction operator.
(b) Diagrams for the matrix elements of the dimer local
operator $\varphi^{\dagger}_{\alpha}(R)\varphi_{\beta}(R)$ and its derivatives $\varphi_{\alpha}^{\dagger}(R)\left[i\partial_{t}+\partial_{R}^{2}/(4m) \right]^{u}(-i\partial_{R})^{v} \varphi_{\beta}^{}(R)$, with $u,v=0,1,2,3,\cdot\cdot\cdot$.
(c) Diagram for the matrix elements of the
operator $\psi_{\sigma}^{\dagger}(R+x) \psi_{\sigma'}(R)$.
(d) Diagram for the matrix element of $\int dt~e^{i\omega t-ikx}\int dx {\cal T}{\cal O}_{\sigma3}(R+x,t){\cal O}_{\sigma'3}^{\dagger}(R,0)$ ($\sigma=\uparrow,\downarrow$).
The single line denotes the atom propagator matrix $G$,
the double lines denote the matrix elements of the dimer propagator matrix $D_{\alpha\beta}$ with
$\alpha,\beta \in \{S,P\}$, the blue dot represents the interaction vertex, $-i{\sigma_{\alpha}}$ or $-i{\sigma_{\beta}}$, and the open dot represents the insertion of operators.} \label{fig:FeynmanDiagrams}
\end{figure}
%%%%%%%%%%%%%%%%%%%%%%%%%%%%%%%%%%%%%%%%%%%%%%%%%%

With the above renormalization relation of $g_P^{}$, the scattering amplitude of the model \eqref{eq:Lagrangian} is finite. Explicitly, the nontrivial part of the scattering amplitude is from the renormalization of the dimer Green's function $D_{\alpha\beta}(E_0,Q)=\left<\varphi_{Q,\alpha}^{}(E_0)\varphi_{Q,\beta}^\dagger(E_0)\right>$, where $E_{0}$ is the total energy. Here the expectation is under the real-time path integral with the Lagrangian \eqref{eq:Lagrangian}. As shown in Fig.~\ref{fig:FeynmanDiagrams}(a), the inverse of the dimer propagator matrix is given by
\begin{align}\label{eq:Dsp}
&~~D^{-1}(E_{0},Q) = \nonumber\\
&\left(\begin{array}{cc}
(ig_{S}^{})^{-1} - \Pi_{SS}(E_{0},Q)  & -\Pi_{SP}(E_{0},Q) \\
-\Pi_{PS}(E_{0},Q)   &  (ig_{P}^{})^{-1} - \Pi_{PP}(E_{0},Q)
\end{array}\right),
\end{align}
where the polarization bubble reads
\begin{align}\label{eq:bubble-SOC}
&\Pi_{\alpha\beta}(E_{0},Q) = -\int\frac{dpd{\omega_{0}}}{(2\pi)^2} \frac{p^{l_{\alpha}+l_{\beta}}}{2} \nonumber\\
&\times\text{Tr}\left[ G^{T}({\omega_{0}},Q/2 + p){\sigma_{\alpha}} G(E_{0}-{\omega_{0}},Q/2 - p){\sigma_{\beta}^{\dagger}} \right],
\end{align}
where $\alpha,\beta \in \{S,P\}$ and we have defined $l_{S}=0$ and $l_{P}=1$. $\text{Tr}$ denotes the trace over the spin degrees of freedom. $G$ is the time-ordered Green's function matrix for fermions defined as $G_{\sigma\sigma'}(\omega,k)=\left<\psi_\sigma(\omega,k)\psi_{\sigma'}^\dagger(\omega,k)\right>$. We have
\begin{align}
[G^{-1}(\omega,k)]_{\sigma\sigma'} =  -i[(\omega+i0^{+})\delta_{\sigma\sigma'} - ({\cal H}_{k}^0)_{\sigma\sigma'}].
\end{align}
The integral in \eqref{eq:Dsp} can be carried out analytically and we present the result with $Q=0$ in the Supplemental Material~\cite{Supp}.
Here, for simplicity, we only present the result for small $k_0$ and $\Omega$:
\begin{equation}\label{eq:expansionbuuble}
D^{-1}(E_0,0)\approx\left(\begin{array}{cc}
-\frac{ma_s}{2}+\frac{m}{2\sqrt{-mE_{0}}}  &\frac{\sqrt{m}k_0\Omega}{8(-E_{0})^{3/2}} \\
\frac{\sqrt{m}k_0\Omega}{8(-E_{0})^{3/2}}  &  \frac{m-a_pm\sqrt{-mE_{0}+k_0^2}}{4a_p}
\end{array}\right).
\end{equation}
We have assumed $E_0<0$ and kept terms up to the $k_0^2$ and $\Omega$ order. The result shows that all divergence can be absorbed by the renormalization relation \eqref{eq:renormalization}. In particular, the off-diagonal terms $\Pi_{SP}$ and $\Pi_{PS}$ are proportional to $k_0\Omega$ and thus finite, indicating the physics is universal. This is due to a nontrivial SOC, where we need both $\Omega$ and $k_0$ to be nonzero. In contrast, for the higher partial-wave systems in higher dimension, additional divergence may appear and new renormalization relations are needed.

\section{Contact matrix}\label{3}
For a dilute atomic gas system described by \eqref{eq:Lagrangian}, we expect universal behaviors governed by two-body physics when we focus on physics at some momentum scale $k$ that satisfies $ \Lambda\gg k\gg \text{max}\{k_F,\sqrt{mT}\}$. Here $k_F$ is the Fermi momentum determined from the density of fermions and $T$ is the temperature.

Theoretically, operator product expansion (OPE) is an ideal tool to explore such universal physics~\cite{Braaten20081,Braaten20082}. One can expand the product of two operators as
\begin{equation}\label{ope}
\left.{\cal O}_i(x+R) {\cal O}_j(R) \right|_{x\rightarrow 0}= \sum_n C_{ij}^k(x) {\cal O}_k(R),
\end{equation}
where $\{{\cal O}_i\}$ is a set of local operators and $C_{ij}^k(x)$ are expansion functions. After the Fourier transform, this gives the major contribution at large momentum. There is a similar expansion in time direction.

For a cold-atom system with only $s-$ or $p-$wave interaction, it is known that the leading-order contribution is from the contact operator {$\hat{C}_{SS}^{(0,0)}(R)$ or $\hat{C}_{PP}^{(0,0)}(R)$, which is given by Eq.~(\ref{eq:contact operator}).} Intuitively, these contact operators count the effective number of dimers in a many-body system. When we turn on SOC, there is a finite correlation between the $s$- and $p$-wave dimers. We expect the system should instead be governed by the contact operator matrix,
{\begin{align}\label{eq:contact operator}
\hat{C}_{\alpha\beta}^{(u,v)}(R) =  m^{2+u}\varphi^{\dag}_{\alpha}(R) \left(i\partial_{t}+\frac{\partial_{R}^{2}}{4m} \right)^{u}(-i\partial_{R})^{v} \varphi_{\beta}^{}(R),
\end{align} where $u,v=0,1,2,3,\cdot\cdot\cdot$. }
The contact matrix of the system is then defined as $C_{\alpha\beta}=\int dR\left<\hat{C}_{\alpha\beta}(R)\right>$. The idea of a matrix form contact was introduced in~\cite{YoshidaTensor2016,ZhangMatrix2017,footnote}.
We now derive the universal relations for the momentum distribution and Raman spectral by matching their asymptotic behaviors with contact operators.

\section{Momentum tail}\label{4}
Physically, we know that SOC should make spin $\uparrow$ and $\downarrow$ different. Hence, we consider the momentum distribution matrix $n_{\sigma'\sigma}(q)=\langle \psi_{q,\sigma}^{\dagger}\psi_{q,\sigma'}\rangle=\int dx dR e^{-iq x}\langle\psi_{\sigma}^{\dag}(R+x) \psi_{\sigma'}(R)\rangle/L$, where $q$ is the relative momentum. This corresponds to considering ${\cal O}_i=\psi_\sigma^\dagger$ and ${\cal O}_j=\psi_{\sigma'}$ in \eqref{ope}.

To determine the coefficient of OPE, we take the matrix elements for both sides of \eqref{ope}. Usually, one considers both incoming and outgoing states with two fermions. However, in our model \eqref{eq:Lagrangian}, two fermions can only interact by first combining to dimers and we could equivalently consider a single incoming dimer $\left|I_{\alpha_{i}}\right\rangle=\int dtdR~e^{i(E_0t-QR)}\varphi_{\alpha_{i}}^\dagger(R,t)\left|0\right\rangle$ and a
single outgoing dimer $\left\langle O_{\alpha_{o}}\right|=\int dt dR~e^{-i(E_0t-QR)}\left\langle 0\right|\varphi_{\alpha_{o}}(R,t)$. Here,  $E_0$ is the total energy and $Q$ is the total momentum.

We first consider the matrix element of the contact operator matrix, which is expected to be the right-hand side of the OPE equation~(\ref{ope}). The corresponding diagram is shown in Fig.~\ref{fig:FeynmanDiagrams}(b):
{\begin{align}\label{eq:Contacts}
\frac{C_{\alpha\beta}^{(u,v)}}{m^{2+u}} &= \int dR\left\langle O_{\alpha_{o}}\right| \varphi^{\dagger}_{\alpha}(R,t) \left(i\partial_{t}+\frac{\partial_{R}^{2}}{4m} \right)^{u} \nonumber\\
&\times(-i\partial_{R})^{v} \varphi_{\beta}^{}(R,t) \left|I_{\alpha_{i}}\right\rangle \nonumber\\
&= \left(E_{0} - \frac{Q^{2}}{4m}\right)^{u}{\bf Q}^{v}D_{\alpha_{o}\alpha}(E_{0},Q) D_{\beta\alpha_{i}}(E_{0},Q),
\end{align} where $E_{0}$ is the total energy and ${\bf Q}$ is the total momentum. Notice that, {in one dimension, the momentum is still a vector because a 1D vector has two opposite directions, and the 1D momentum can be defined as~\cite{Barlette_2000} ${\bf q}\equiv|q|\text{sgn}(q)$ with the signum function $\text{sgn}(q)$. Therefore, the quantity ``bold Q" is defined as ${\bf Q}\equiv|Q|\text{sgn}(Q)$ which means that the 1D center-of-mass momentum has two opposite directions. In this case, a vector or a scalar can be distinguished by their representation under inversion.} If $v$ is an odd number, the corresponding contact is a vector. This is to be matched with the matrix element of $\psi_{\sigma}^{\dag}(R+x) \psi_{\sigma'}(R)$. The non-trivial interaction effect comes from the diagram shown in Fig.~\ref{fig:FeynmanDiagrams}(c).
After the Fourier transform, we get the momentum distribution matrix as
\begin{align}\label{eq:distribution}
&n(q)
= \sum_{\alpha,\beta=S,P}(-i)^{2}D_{\alpha_{o}\alpha}(E_{0},Q)D_{\beta\alpha_{i}}(E_{0},Q)\int_{-\infty}^{\infty} \frac{d{\omega_{0}}}{2\pi} \nonumber\\
&\times q^{l_{\alpha}+l_{\beta}}G(E_{0}-{\omega_{0}},q) {\sigma_{\beta}} G^{T}({\omega_{0}},Q-q){\sigma_{\alpha}^{\dagger}} G(E_{0}-{\omega_{0}},q).
\end{align}
Keeping every element up to the order in the $1/q^{4}$ expansion, we
have the momentum distribution matrix,
{\begin{widetext}\begin{eqnarray}\label{eq:distribution-q}
n(q)
\sim
\left(\begin{array}{cc}
\frac{C_{PP}}{q^{2}L} + \frac{2\hat{\bf q}\cdot{\bf C}_{Q1}}{q^{3}L} + \frac{2C_{r}-2k_{0}^{2}C_{PP}-2k_{0}\hat{\bf q}\cdot{\bf C}_{Q1}+5C_{Q2}/2}{q^{4}L} + \frac{C_{SS}}{q^{4}L}  & -\frac{C_{SP}}{q^{3}L} - \frac{2k_{0}C_{SP}+2\hat{\bf q}\cdot{\bf C}_{SPQ1}}{q^{4}L} - \frac{m\Omega C_{PP}}{q^{4}L} \\
-\frac{C_{PS}}{q^{3}L} - \frac{2k_{0}C_{PS}+2\hat{\bf q}\cdot{\bf C}_{PSQ1}}{q^{4}L} - \frac{m\Omega C_{PP}}{q^{4}L}  &  \frac{C_{SS}}{q^{4}L}
\end{array}\right),
\end{eqnarray}\end{widetext} where {``$ \sim $" means expanding to a certain order in the large-$q$ limit, $\hat{\bf q}\equiv{\bf q}/|q|$} is the unit vector, and we use $C_{PP}=C_{PP}^{(0,0)}$, ${\bf C}_{Q1}=C_{PP}^{(0,1)}$, $C_{r}=C_{PP}^{(1,0)}$, $C_{Q2}=C_{PP}^{(0,2)}$, $C_{PS}=C_{PS}^{(0,0)}$, $C_{SP}=C_{SP}^{(0,0)}$, ${\bf C}_{PSQ1}=C_{PS}^{(0,1)}$, ${\bf C}_{SPQ1}=C_{SP}^{(0,1)}$, and $C_{SS}=C_{SS}^{(0,0)}$.}
Recall that the effective Lagrangian \eqref{eq:Lagrangian} is different from that in the laboratory frame by a momentum shift. For subleading terms, this momentum shift would modify the coefficient, as in \cite{ZhangSOC2018,QinSOC20201}. However, the leading-order results in Eq.~\eqref{eq:distribution-q} are free from such complications. Moreover, note that this derivation can also be carried out for systems without SOC, which leads to the same leading-order results in Eq.~\eqref{eq:distribution-q}. However, in that case, we have $C_{SP}=C_{PS}=0$ due to the reflection symmetry. Here, the SOC plays a role of breaking the rotational symmetry and making $C_{SP}$ and $C_{PS}$ finite.

Experimentally, we could measure each component separately and extract their leading-order behaviors. As an example, for the off-diagonal terms, we could measure the momentum of fermions in the spin states $\left|\pm x\right\rangle=\frac{1}{\sqrt{2}}(\left|\uparrow\right\rangle\pm\left|\downarrow\right\rangle)$. {Up to the leading order, this gives}$$n_{++}(q)-n_{--}(q)=n_{\downarrow\uparrow}(q)+n_{\uparrow\downarrow}(q)\sim -\frac{C_{PS}+C_{SP}}{q^{3}L}.$$ Similarly, measuring in the spin states $\left|\pm y\right\rangle$ gives $C_{PS}-C_{SP}$.

\section{Raman spectroscopy}\label{5}
The Raman spectroscopy can be used as an important experimental tool in cold atom systems. When the transfer momentum and frequency are large, the Raman spectroscopy can also be related to the contacts. We consider applying a Raman coupling with frequency $\omega>0$ and momentum $k$ to transfers fermions from the internal spin state $|\sigma\rangle$ ($\sigma=\uparrow,\downarrow$) into a third spin state $|3\rangle$. The Hamiltonian reads
$H_c=\sum_\sigma\Omega_{\sigma}\int dx ~e^{i(kx-\omega t)}{\cal O}_{\sigma3}(x,t)+\text{H.c.},$ where ${\cal O}_{\sigma3}(x,t) \equiv \psi_{3}^{\dag}(x,t)\psi_{\sigma}(x,t)$. The transition rate function $R(\omega,k)$ to $|3\rangle$ is given by the Fermi golden rule, which is related to the imaginary part of the time-ordered two-point correlation function~\cite{BraatenRF2010,HofmannRF2011}:
\begin{align}\label{eq:rf}
R(\omega,k)&=2\pi\sum_{\sigma\sigma'}\Omega_\sigma\Omega_{\sigma'}^*\Gamma^{R}_{\sigma\sigma'}(\omega,k) {,} \\
\Gamma^{R}_{\sigma\sigma'}(\omega,k) &= \frac{1}{\pi} \text{Im}~\int dR \int dt~e^{i\omega t} \int dx~e^{-ikx}\nonumber\\
&\times i\left\langle {\cal T}{\cal O}_{\sigma3}(R+x,t){\cal O}_{\sigma'3}^{\dag}(R,0) \right\rangle,
\end{align}
where ${\cal T}$ is the time-ordering operator. We thus study the OPE of ${\cal O}_{\sigma3}$ and ${\cal O}_{\sigma'3}^{\dag}$. The diagram is shown in Fig.~\ref{fig:FeynmanDiagrams}(d):
\begin{equation}
\begin{aligned}\label{eq:rf-SOC}
&\Gamma^{R}_{\sigma\sigma'}(\omega,k)\\ =& \frac{1}{\pi}\text{Im}~i\sum_{\alpha,\beta=S,P}(-i)^{2}D_{\alpha_{o}\alpha}(E_{0},Q)D_{\beta\alpha_{i}}(E_{0},Q) \\
&\times\int \frac{dp d{\omega_{0}}}{(2\pi)^{2}} p^{l_{\alpha}+l_{\beta}}G_0(E_0-{\omega_{0}}+\omega,p+k)  \\
&\times\left[G(E_{0}-{\omega_{0}},p){\sigma_{\beta}^{\dagger}} G^{T}({\omega_{0}},Q-p){\sigma_{\alpha}} G(E_{0}-{\omega_{0}},p)\right]_{\sigma\sigma'}.
\end{aligned}
\end{equation}
Matching Eq.~(\ref{eq:rf-SOC}) with Eq.~(\ref{eq:Contacts}), we have the Raman transfer rate in the high-frequency and large-momentum limit,
\begin{equation}
\begin{aligned}
&\Gamma^{R}(\omega,k) =\frac{2m}{\pi\sqrt{4 m\omega -k^2}}\times \\
&~~~\left(\begin{array}{cc}
\frac{2 m\omega C_{PP}}{\left(k^2-2 m\omega \right)^2 }  & \frac{k\left(k^{2}-6m\omega \right)C_{SP}}{\left(k^{2}-2m\omega \right)^{3} } \\
\frac{k\left(k^{2}-6m\omega \right)C_{PS}}{\left(k^{2}-2m\omega \right)^{3} }  &  \frac{2\left[4 (m\omega)^2+4 k^2 m\omega -k^4\right]C_{SS}}{\left(k^2-2 m\omega \right)^{4}
  }
\end{array}\right).
\end{aligned}
\end{equation}
Here we have assumed $\omega>k^2/(4m)$ and kept each element to the leading order. Taking the limit of $k=0$ leads to the high-frequency tail of the radio-frequency spectral $\Gamma^{rf}_{\sigma\sigma'}(\omega)=\Gamma^{R}_{\sigma\sigma'}(\omega,0)$,
{\begin{align}
\Gamma^{rf}(\omega) &=  \frac{m}{2\pi}
\left(\begin{array}{cc}
\frac{C_{PP}}{(m\omega)^{3/2}}  & 0 \\
0  &  \frac{C_{SS}}{(m\omega)^{5/2}}
\end{array}\right).
\end{align}

The result of $\Gamma^{R}(\omega,k)$} provides an individual experimental observable to determine different contacts by tuning $\Omega_\sigma$ \eqref{eq:rf}. The Raman spectroscopy, together with the momentum distribution, serves as a nontrivial check for the universal relations in the hybridized system \eqref{eq:Lagrangian}.

\section{Other universal relations}\label{6}
In this section, we discuss other universal relations, including the adiabatic relations and thermodynamical relations. Since the derivation is standard, we focus on presenting the results here and we give details of the derivations in the Supplemental Material~\cite{Supp}.

We first focus on the adiabatic relations. The traditional $s/p$-wave contacts correspond to the change of energy when varying $a_s$ or $-1/a_p$, which can be seen from taking the derivative with $g_\alpha$ in the Lagrangian \eqref{eq:Lagrangian} as
\begin{align}
\frac{C_{SS}}{2m} &\equiv \frac{\partial E}{\partial a_{s}} ,\ \ \ \ \ \
\frac{C_{PP}}{4m} \equiv - \frac{\partial E}{\partial a_{p}^{-1}}. \label{eq:cpE}
\end{align}
However, there is no direct $s-$ and $p-$wave dimer mixing in \eqref{eq:Lagrangian} and thus no adiabatic relation for $C_{SP}$ or $C_{PS}$. On the other hand, we could consider a nonspherical potential between atoms where microscopic mixing terms $\delta_{SP}^{}\varphi^{\dagger}_{Q,S}\varphi_{Q,P}^{}+\text{H.c.}$ exist in the action. In this case, the off-diagonal components of the contact matrix correspond to varying $\delta_{SP}$.

When SOC is present, there are two new parameters $k_{0}$ and $\Omega$.
One can define two new contacts $C_{\lambda}$ and $C_{\Omega}$ as
\begin{align}
C_{\lambda} &\equiv  \frac{\partial E}{\partial k_0} , \ \ \ \ \ C_{\Omega} \equiv \frac{\partial E}{\partial \Omega}.\label{eq:cOmega}
\end{align}
Here, $C_{\lambda}$ and $C_{\Omega}$ only refer to single-atom operators which give nonzero matrix elements
in the single-atom sector. The momentum distribution under
single-particle states is just a delta
function, so that $C_{\lambda}$ and $C_{\Omega}$ will not
contribute to the large-momentum tail, which is different from $C_{SS}$ and $C_{PP}$. However, both $k_{0}$ and $\Omega$ have a nonzero energy scale, so that they would appear in the pressure relation and viral theorem. For a uniform gas, the pressure relation reads
\begin{align}
{\cal P} = 2{\cal E} + \frac{a_{s}C_{SS}}{2mL} + \frac{C_{PP}}{4ma_{p}L} - \frac{k_{0}C_{\lambda}}{L} - \frac{2\Omega C_{\Omega}}{L}, \label{eq:pressure}
\end{align} where ${\cal E}=E/L$ is the energy density. For an atomic gas in a harmonic potential $V_T=m\omega^2 x^2/2$ with the trapping frequency $\omega$, the viral theorem is written as:
\begin{equation}
E = 2 \langle V_T\rangle - \frac{a_{s}C_{SS}}{4m } - \frac{C_{PP}}{8ma_{p}} + \frac{k_{0}C_{\lambda}}{2} + \Omega C_{\Omega}  \label{eq:virial}
\end{equation}
with $\langle V_T\rangle$ being the trapping energy.

%%%%%%%%%%%%%%%%%%%%%%%%%%%%%%%%%%%%%%%%%%%%%%%%%%
\begin{figure}
\includegraphics[width=8.8cm]{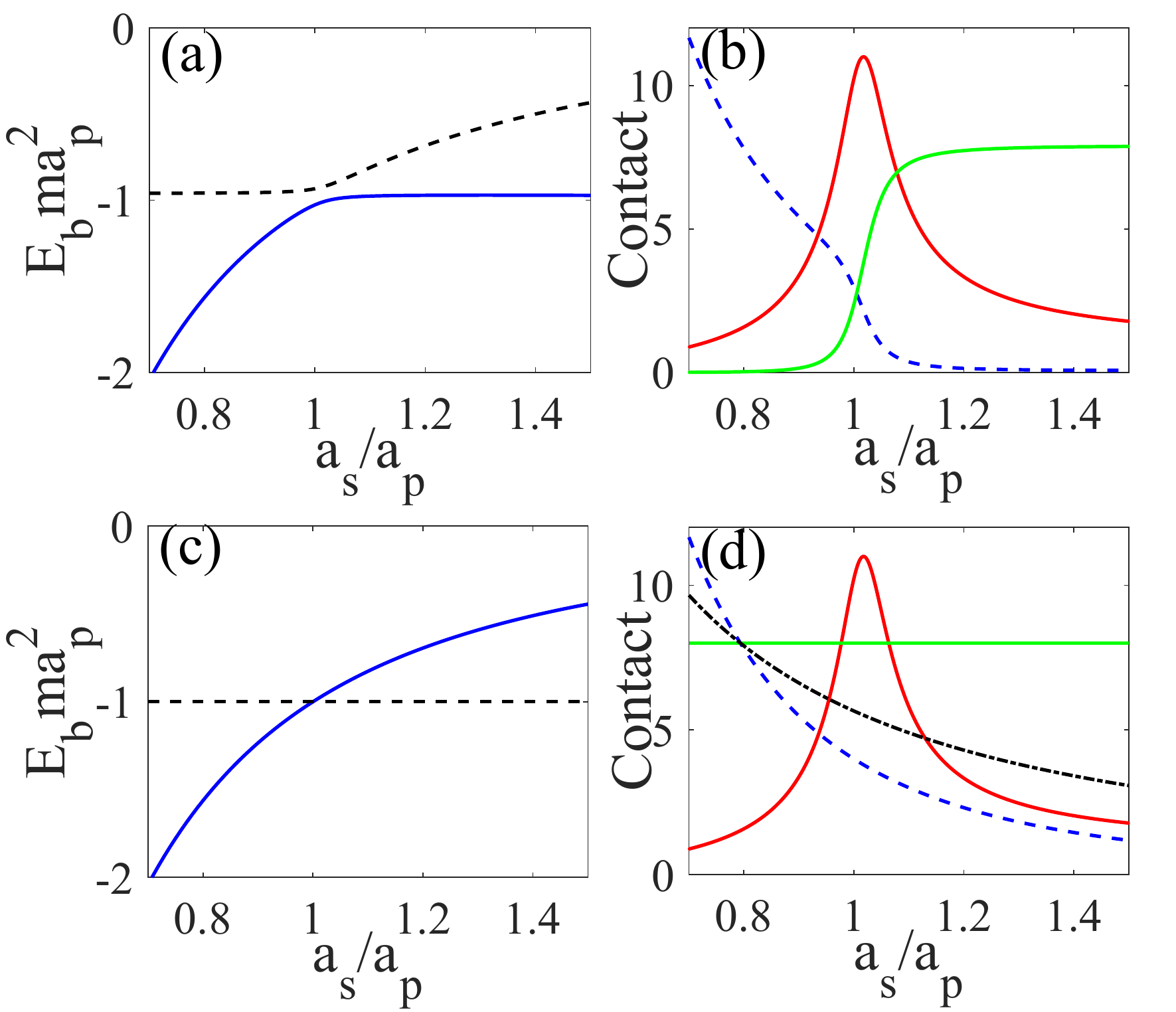}
\caption{{(a) Dimensionless two-body banding energy vs $a_{s}/a_{p}$ with SOC. The black dashed curve denotes $E_{b}^{(+)}ma_{p}^{2}$ and the blue solid curve denotes $E_{b}^{(-)}ma_{p}^{2}$. (b) Dimensionless two-body contacts vs $a_{s}/a_{p}$ with SOC. (c) Dimensionless two-body banding energy vs $a_{s}/a_{p}$ without SOC. The black dashed curve denotes $E_{b}^{(p)}ma_{p}^{2}$ and the blue solid curve denotes $E_{b}^{(s)}ma_{p}^{2}$. (d) Dimensionless two-body contacts $C_{SS}$ and $C_{PP}$ vs $a_{s}/a_{p}$ without SOC. As a comparison, we also plot the $C_{SP}$ with finite SOC [the same curve as (b)]. } The red solid curve denotes $C_{SP}a_{p}^{2}$, the blue dashed curve denotes $C_{SS}a_{p}^{3}$, the green solid curve denotes $C_{PP}a_{p}$, and {the black dot-dashed curve denotes $\sqrt{C_{SS}C_{PP}}a_{p}^{2}$.} Here, we choose the SOC parameters as $k_{0}a_{p}=0.2$ and $m\Omega a_{p}^{2}=0.3$. \label{fig:TwobodyContacts}}
\end{figure}
%%%%%%%%%%%%%%%%%%%%%%%%%%%%%%%%%%%%%%%%%%%%%%%%%%

\section{Contacts in two-body bound states}\label{7}
To give an explicit example of the contact matrix in the hybridized $s$- and $p$-wave system, we now perform a calculation for the two-body bound state. Generally, the binding energy $E_b$ with momentum $Q$ is given by solving $\text{det}(D^{-1}(E_0,Q))=0$. We consider the case with small SOC strength where we could use \eqref{eq:expansionbuuble}.

We focus on $Q=0$ with both $a_s>0$ and $a_p>0$. For $\Omega=0$, there is both an $s$-wave bound state with binding energy $E_b^{(s)}=-1/(ma_s^2)$ and a $p$-wave bound state with binding energy $E_b^{(p)}=-1/(ma_p^2) + k_0^2/m$. Here the presence of $k_0$ is because $Q=0$ corresponds to a center-of-mass momentum $2k_0$ for the $p$-wave bound state in the laboratory frame. {In this case, we have $C_{SS}=4/a_{s}^{3}$, $C_{PP}=8/a_p$, and $C_{SP}=C_{PS}=0$.}

When we turn on finite but small $\Omega$, the binding energies receive an important correction only near the resonance with $1/(a_s^0)^2=1/(a_p^0)^2-k_0^2.$ We then approximate
\begin{equation}
D^{-1}(E_{b},0)\approx
\left(\begin{array}{cc}
I_1\left(E_{b}-E_b^{(s)}\right)  & K_\Omega \\
K_\Omega  &  I_2\left(E_{b}-E_b^{(p)}\right)
\end{array}\right),
\end{equation}
where $I_1=\frac{m^{2}a_s^3}{4}$, $I_2=\frac{m^{2}a_{p}}{8}$, and $K_\Omega=\frac{k_0\Omega m^{2}(a_s^0)^3}{8}$. Then the binding energy can be derived as
\begin{equation}
\begin{aligned}
2E_{b}^{(\pm)}=&E_b^{(p)}+E_b^{(s)} {\pm}
\\&\sqrt{(E_b^{(p)})^2-2E_b^{(p)}E_b^{(s)}+(E_b^{(s)})^2+\frac{4K_\Omega^2}{I_1I_2}}.
\end{aligned}
\end{equation}
The contacts $C_{SS}$ and $C_{PP}$ can be derived by taking the derivation with $a_s$ or $-1/a_p$. To calculate $C_{SP}$ or $C_{PS}$, we apply the trick by adding the additional $\delta_{SP}$ terms and set them to be zero after taking derivatives.

The explicit formula for all contacts are given in the Supplemental Material~\cite{Supp}. A plot for $E_{b}^{(\pm)}$ and contacts for $E_{b}^{(-)}$ are shown in Figs. \ref{fig:TwobodyContacts}(a) and \ref{fig:TwobodyContacts}(c). Away from the degenerate point, $E_{b}^{(\pm)}$ approaches $E_b^{(s)}$ or $E_b^{(p)}$.
Comparing Figs. \ref{fig:TwobodyContacts}(a) with \ref{fig:TwobodyContacts}(c), it is found that the SOC parameters can open a gap between the two banding energies $E_{b}^{(+)}$ and $E_{b}^{(-)}$.
Consequently, for the diagonal components of the contact matrix, we have $C_{SS}\approx 0$ for $a_s/a_p\gg 1$ and $C_{PP}\approx 0$ for $a_s/a_p\ll 1$. Near the degenerate point $a_s/a_p\sim 1$, we see a peak for $C_{SP}$, indicating a large mixing between $s$- and $p$-wave dimers as expected.
Moreover, we also calculate the amplitude of the hybridized new contacts compared to the $s$- and $p$-wave ones without SOC, as shown in Fig. \ref{fig:TwobodyContacts}(d), to give the possibility of the measurement.

\section{Discussions}\label{8}
In this work, we have derived the momentum tail and the Raman spectroscopy for hybridized $s$- and $p$-wave interactions from spin-orbital coupling in 1D. We find new contacts appear at the leading order of certain observables due to the mixing between different partial waves.

We finally comment on the generalization to higher-dimensional systems with 1D (NIST) SOC. In higher dimensions, first, we have the additional quantum number $m=-1,0,1$ in 3D or $m=\pm 1$ in 2D for $p$-wave dimers. Depending on whether their resonance splits, we may have a larger contact matrix. To the leading order, the off-diagonal components of the momentum distribution should again correspond to the off-diagonal contacts and should be proportional to $1/q^3$. On the contrary, the scaling of the Raman spectral would change (by a factor of $\sim \omega^{(D-1)/2}$ for large $\omega$) due to the difference of the density of state.

\begin{acknowledgments}
We thank Xiaoling Cui and Shi-Guo Peng for helpful discussions.
This work is supported by the National Natural Science Foundation of China (Grant No. 11404106).
F.Q. acknowledges support from the project funded by the China Postdoctoral Science Foundation (Grants No. 2019M662150 and No. 2020T130635) and the SUSTech Presidential Postdoctoral Fellowship.
\end{acknowledgments}

%\bibliography{refs-1dsoc}
%

\end{document}